\documentclass[twocolumn]{article}

\usepackage[utf8]{inputenc}
\usepackage{authblk}
\usepackage{natbib}
\usepackage{url}
\usepackage{graphicx}

\title{VIRUP : The Virtual Reality Universe Project}

\author[1]{Florian Cabot}
\author[1]{Yves Revaz}
\author[1]{Jean-Paul Kneib}
\author[2]{Hadrien Gurnel}
\author[2]{Sarah Kenderdine}

\affil[1]{Institute of Physics, Laboratoire d'Astrophysique, \'{E}cole Polytechnique F\'{e}d\'{e}rale de Lausanne, CH-1015 Lausanne, Switzerland}

\affil[2]{Laboratory for Experimental Museology, \'{E}cole Polytechnique F\'{e}d\'{e}rale de Lausanne, CH-1015 Lausanne, Switzerland}

\date{}                     

\begin{document}

\maketitle

\begin{abstract}
\texttt{VIRUP}\footnote{\url{http://go.epfl.ch/virup}} is a new C++ open source software that provides an interactive virtual reality environment to navigate through large scientific astrophysical datasets 
obtained from both observations and simulations.
It is tailored to visualize terabytes of data, rendering at 90 frames per second in order to ensure an optimal immersion experience.
While \texttt{VIRUP} has initially been designed to work with gaming virtual reality headsets, it supports
different modern immersive systems like 3D screens, $180^{\circ}$ domes or $360^{\circ}$ panorama.
\texttt{VIRUP} is scriptable thanks to the \texttt{Python} language, a feature that allows to 
immerse visitors through pre-selected scenes or to  pre-render sequences to 
create movies. 
A companion video 
\footnote{\url{https://www.youtube.com/watch?v=KJJXbcf8kxA}}
to the last SDSS 2020 release as well as a 
21 minute long documentary, \emph{The Archaeology of Light} 
\footnote{\url{https://go.epfl.ch/ArchaeologyofLight}}
have been both 100\% produced using \texttt{VIRUP}.  


\end{abstract}


\section{Goals}


The goal of 
\texttt{VIRUP}, the Virtual Reality Universe Project is to provide the most comprehensive 
view of our Universe through the most modern visualisation techniques : Virtual Reality (VR). 
By combining observational and computational astrophysics data, we aim at providing 
a tool to access the most modern 3D comprehensive view of our Universe, from cosmological scales down to artificial 
satellites orbiting the Earth.
We aim to let anyone understand the hierarchical organization of our universe and to help
develop intuitions for astrophysical processes that shaped the Universe and its content 
as we observe it now.

\texttt{VIRUP}  is concretised by the creation of a new C++/OpenGL/Qt
flexible free software of the same name, built on top of a custom-designed graphics engine. 
While quite demanding, developing a brand new graphics engine was necessary 
to reach optimal performances, in particular when displaying very large data sets.  
It was also required to avoid any constraint imposed by standard existing engines.

In this paper, we give a very brief overview of the features currently implemented in 
\texttt{VIRUP}.
A forthcoming paper will describe in more detail the numerical techniques that were used.


\section{Features}


In its default mode, \texttt{VIRUP} lets the user navigate interactively through the 
different scales of the Universe. An immersive experiment is
obtained if a virtual reality headset is used. In this case, 
the navigation is performed using standard hand-tracking systems
and allows for example to zoom over a chosen object.

\texttt{VIRUP} is supplemented with a semi-interactive mode, in which scenes can be selected by an operator or from a running script. This mode assists the user during the journey in the virtual universe, by guiding them towards the most interesting scenes.
This mode avoids the use of the hand-tracking system which can sometimes require a 
time consuming and overwhelming training, but lets the visitor benefit from a full immersive environment in which they can still roam. This mode is especially designed to offer a quick
initiation to a group of people.

At the Solar System scale, time can be set arbitrarily to focus, for example, on a specific
planet configuration, on a solar eclipse or on the peculiar position of a space probe or artificial satellite.

The python scripting extension of \texttt{VIRUP} makes it possible to manipulate most of the objects of
the rendering engine. This allows to fully script complex sequences including
transitions between them.
This feature has been used to create movies that can be displayed in 2D, 3D, $360^{\circ}$ or
3D-VR-$360^{\circ}$. 
A companion video to the eBoss press release has been created in 2020
\footnote{\url{https://www.youtube.com/watch?v=KJJXbcf8kxA}}.
Recently, we produced a 21 minute long documentary, \emph{The Archaeology of Light},
a journey through the different scales of our Universe
\footnote{\url{https://go.epfl.ch/ArchaeologyofLight}}.


\section{Overview of some viewable data sets}


Data visualized in \texttt{VIRUP} can be taken from both simulations and observations. 
For now, \texttt{VIRUP} has been tested to visualize data from over 8 databases. 
Extending the visualization with more data from other datasets is easy, thanks to the use of CSV, JSON and HDF5 formats as input  to \texttt{VIRUP} and its tools.

\begin{figure}[h]
  \centering
  \includegraphics[width=0.48\textwidth]{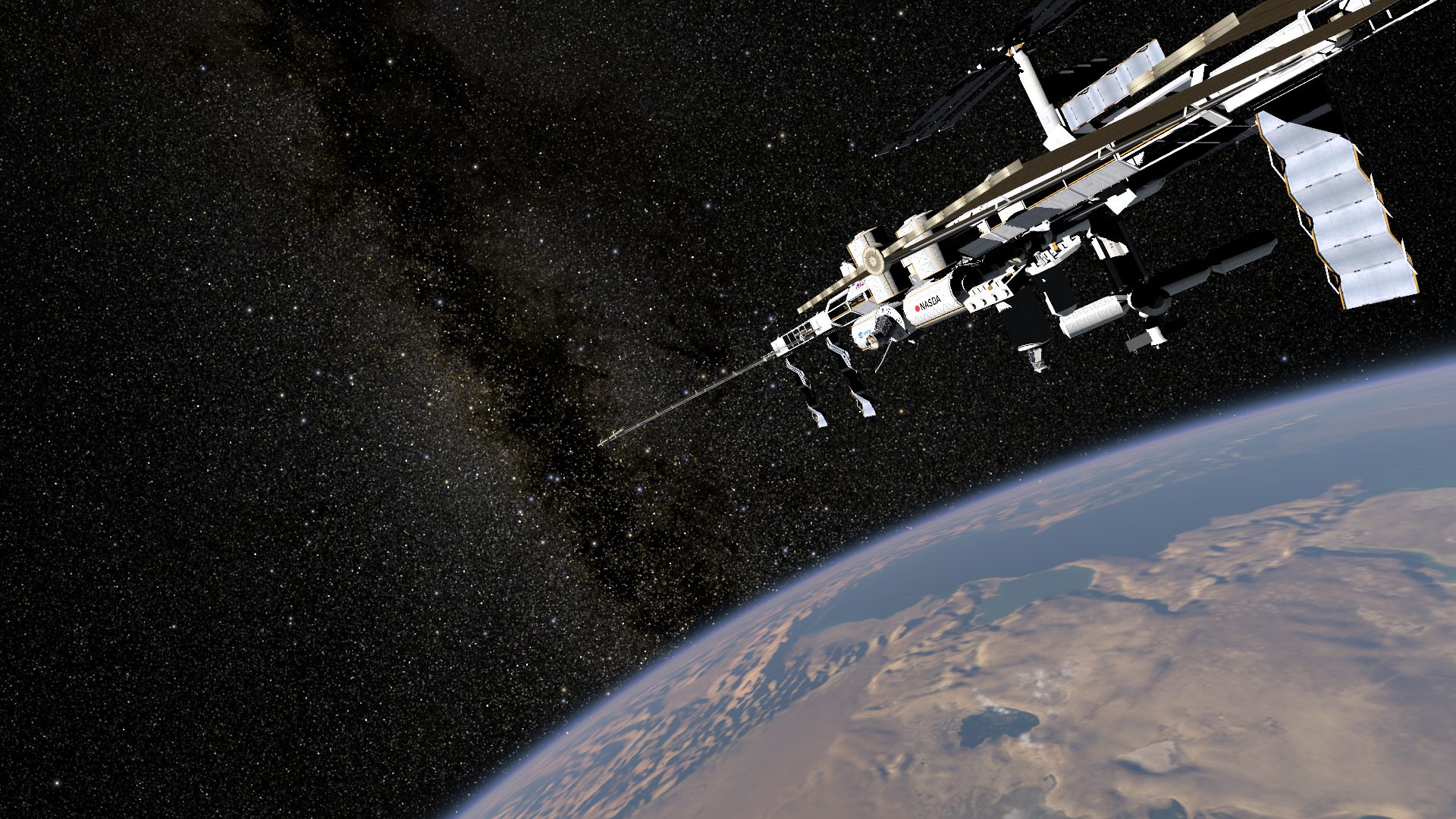}
  \caption{The International Space Station orbiting above the earth. The Milky Way on the background is reproduced by the Gaia data.}

\end{figure}

At the Solar System scale, \texttt{VIRUP} includes both natural and artificial objects. Among them, 
a catalogue of over 3000 satellites orbiting the Earth and other spacecrafts, such as the Voyager 
and Pioneer probes. The Solar System itself is rendered for a particular date and time of day thanks to the data extracted from the tool of NASA JPL Horizons \footnote{\url{https://ssd.jpl.nasa.gov/horizons.cgi}}. The appearance of the bodies is based on data from various NASA missions, extrapolated artistically if necessary. \footnote{Full credits for artistic extrapolations are found in the sub-directories of : \url{https://gitlab.com/Dexter9313/prograde-data}} 3D models of various bodies have also been obtained through the 3D Asteroid Catalogue\footnote{Greg Frieger:\url{https://3d-asteroids.space/}}.

The data for the stars are taken from the Hipparcos and/or the Gaia catalogue (Gaia EDR2), which contains 1.5 billion light sources. 
More than 4500 Exoplanets are included via the Open Exoplanet Catalog \footnote{\url{http://www.openexoplanetcatalogue.com/}}, which aggregates various sources of exoplanet data.

The Milky Way is represented using a realistic high-resolution simulation with initial conditions taken
from the AGORA High-resolution Galaxy Simulations Comparison Project \citep{kim2016} and simulated with 
\texttt{GEAR} \citep{revaz2012}.
Local group dwarf galaxies are included through the continuously-updated
\emph{Local Group and Nearby Dwarf Galaxies} database
of \citet{McConnachie_2012}
\footnote{\url{http://www.astro.uvic.ca/~alan/Nearby_Dwarf_Database.html}}.

To represent the observed large scales of the Universe, \texttt{VIRUP} includes the
Sloan Digital Sky Survey data (SDSS DR16) that consists of over 3.5 million objects \citep{ahumada2020}. 
Simulated portions of the Universe are taken either from the Eagle project \citep{schaye2015}, 
covering a volume of $(100\,\rm{Mpc}/h)^3$ with 6.6 billion particles, 
or the IllustrisTNG project \citep{pillepich2018,springel2018}, with a volume of $(205\,\rm{Mpc}/h)^3$ and 30 billion particles.

Finally, \texttt{VIRUP} also contains the Cosmic Microwave Background radiation map as observed by the Planck Mission
\citep{planck2020} which is used to illustrate the size of the observable Universe.

%
%




\section{Multi-modal visualisation systems}\label{systems}


In addition to standard VR systems like gaming virtual reality headsets,
\texttt{VIRUP} also supports standard 3D screens and has been adapted
to run with more advanced immersion systems, including
$180^{\circ}$ domes (2D or 3D), panoramic or CAVE projection screens,
which combine the images delivered by several projectors.

\texttt{VIRUP} has been tested so far on the following systems 
of the Laboratory for Experimental Museology (eM+)
\footnote{\url{https://www.epfl.ch/labs/emplus/}}: the Cupola, 0.5 CAVE and Panorama+.

\paragraph{The Cupola} is a 5m diameter $\times$ 6m high visualisation system featuring two 3D beamers, whose images are projected onto a hemispherical screen with an effective resolution of $4096\times 4096$ pixels. The screen is positioned above visitors, to allow for a unique viewing experience in a lying position. Since its conception in 2004, this visualisation system was featured in multiple worldwide exhibitions, including \textit{Look Up Mumbai} (2015), \textit{Inside the Ethereal Eye} (2016) and \textit{Seven Sisters} (2017), but also in miscellaneous installations at the UNSW Art \& Design EPICentre (2016-2019).

\begin{figure}[h]
  \centering
  \includegraphics[width=0.48\textwidth]{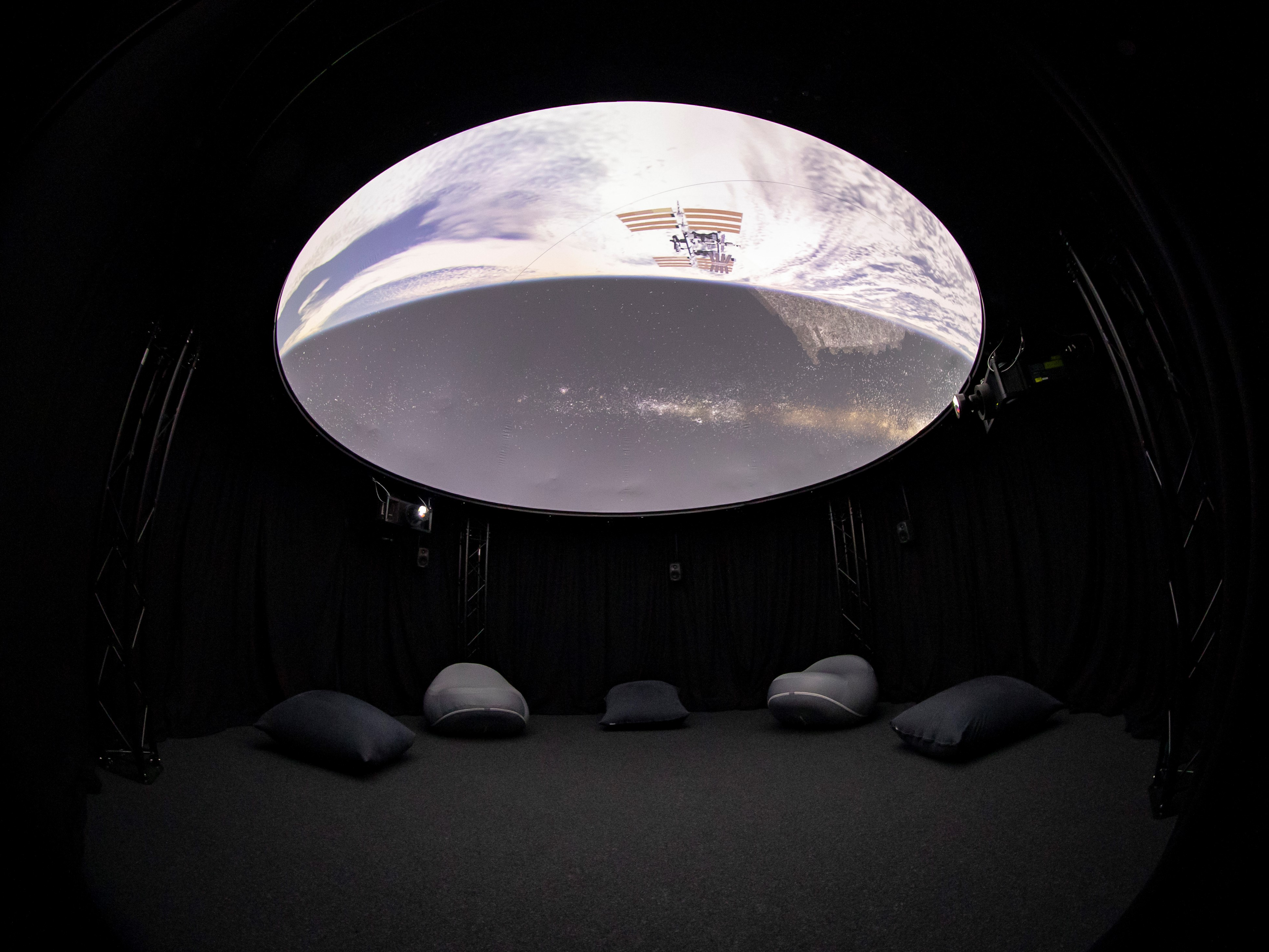}
  \caption{The Earth and International Space Station displayed by the Cupola at the eM+ laboratory}
\end{figure}

\paragraph{The 0.5 CAVE} is a structural modification of the original CAVE visualisation system. Instead of projecting
onto three walls and the floor, it displays images on a single wall and the floor, thanks to two $2560\times 1600$ 3D beamers. One reason for this modification was the need to create a simplified touring model of this installation. It also offers a major benefit, \i.e. its open-viewing configuration, which allows for a much larger public to engage with the displayed content. The 0.5 CAVE uses stereo projection across the wall and floor, making it an ideal immersive space for a wide range of contents.

\begin{figure}[h]
  \centering
  \includegraphics[width=0.48\textwidth]{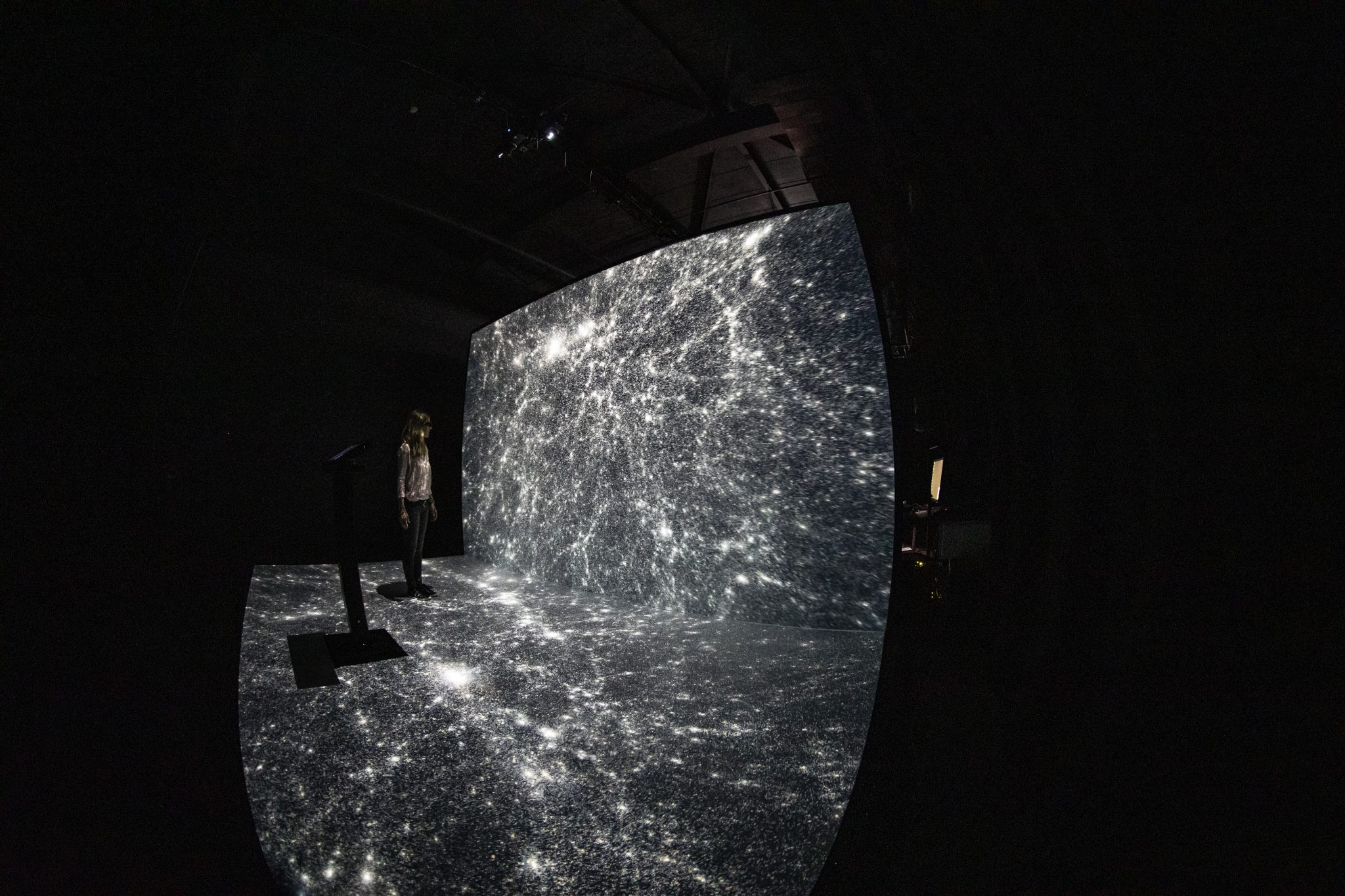}
  \caption{The large scale structure of the Universe from the IllustrisTNG simulations displayed by the 0.5 CAVE at the eM+ laboratory}
\end{figure}

\paragraph{The Panorama+} is a 9m diameter $\times$ 3.6m high 360-degree stereoscopic system, also known as the Advanced Visualisation and Interaction Environment (AVIE). It has an effective resolution of $18000\times 2160$ pixels, which is obtained by combining the images of a cluster of ten computers. Those images are then projected on the screen by five 3D beamers. This system makes it possible for visitors to explore wide virtual environments, as well as large datasets. The first AVIE was launched at UNSW in 2006 and since then, multiple subsequent systems were deployed across the world.      

\begin{figure}[h]
  \centering
  \includegraphics[width=0.48\textwidth]{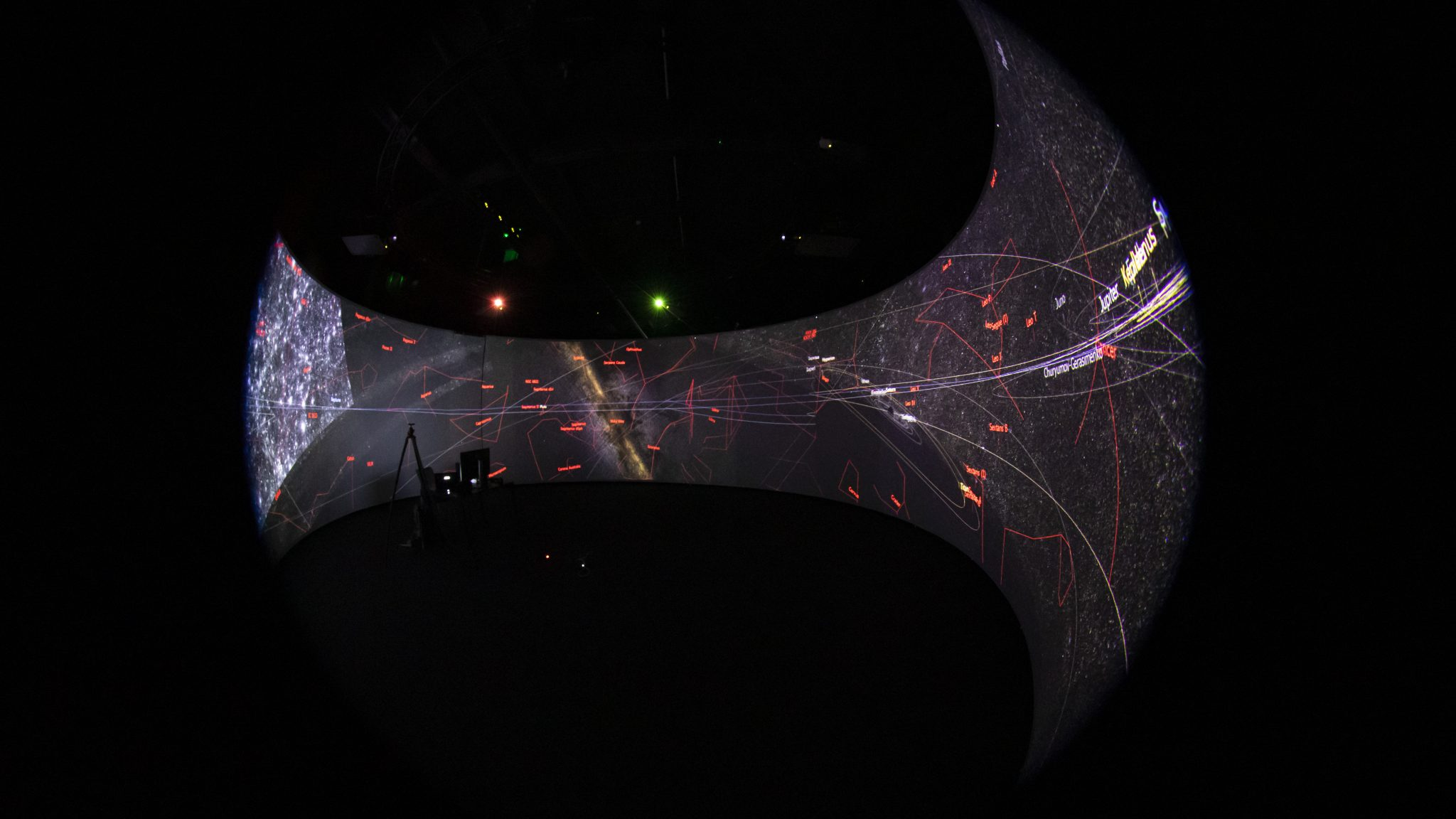}
  \caption{The \texttt{VIRUP} sky displayed by the Panorama+ at the eM+ laboratory}
\end{figure}
%


\section{Challenges and Algorithms}


The most important challenge faced when developing \texttt{VIRUP} has been
to load and render the large and varied datasets on the Graphics Processing Units (GPU), 
ensuring a frame rate of 90 images per second, essential to avoid sickness problems common
in virtual reality environments and guarantee a fully immersive 
and smooth experience. 
A specific effort has been made to render several billions of points contained in 
cosmological simulations.
Ensuring a smooth transition from one database to the next, without freezing the system was
essential.

Another challenge was to represent object over spacial scales spanning up to 27 orders of magnitude, 
from the meter, the typical scale of an artificial satellite, to about $100\,\rm{Gly}$,
the size of the observable Universe. This is especially challenging when working with single-precision variables, which are suited to rendering on GPUs for performance reasons.

Finally, porting \texttt{VIRUP} towards the multi-projector systems described in Sec.~\ref{systems}  required to properly 
synchronize several \texttt{VIRUP} instances running in parallel on different computers.

\subsection{Cosmological Large Scale structures}

\subsubsection{Octree}

As simulation datasets are often very large (several billions of particles), they cannot be rendered 
at once on standard GPUs at an acceptable frame rate required by VR systems. 
To solve this problem, data must be split into chunks,
in order to render only what is currently seen by the user with an acceptable level of detail. The
technique we implemented is greatly inspired from 
\citet{szalay2008}. We construct a LOD-octree by splitting any node (portion of the space) that contains more than 16'000 particles by 8. The octree generated is closer to a kd-tree to ensure node balancing, i.e. we choose a splitting point along all three dimensions to ensure an equal number of particles in each sub-node.

During the rendering process, while walking the tree, we use a rough approximate open angle criterion (taken as the diameter of the node divided by its distance to the camera) to determine if recursion through a node is needed or not. A node that appears large on screen has to render its sub-nodes. On the contrary, a node that appears small is rendered alone without considering its
sub-nodes (a node contains a sample of the content of its children nodes). 
A leaf node will always be rendered if asked to do so, as it has no children.

In order to determine what is physically large enough to trigger a recursion, we could in principle 
setup the opening criterion to a constant value that would fit most situations well. 
However, before rendering, we have no information on the 
shape of the data to be displayed,
on the region that will be observed by the user, nor
on the type of hardware used. 
To avoid any surpise that could lead to a jerky rendering or a freeze-out, 
we developed a flexible solution relying on the current frame rate. 
We use a PID-loop (proportional, integral, derivative) to control the open angle criterion to maintain a given frame rendering time at any time.
For a comfortable VR experience it is set to $11\,\rm{ms}$, i.e. $90\,\rm{FPS}$.

\paragraph{\texttt{octreegen}}
To ensure the fastest possible loading of nodes, the octree is generated offline with a separate tool
called \texttt{octreegen} \footnote{\url{https://gitlab.com/Dexter9313/octree-file-format/}}. The resulting octree is encoded in a specifically tailored file format which can be read by \texttt{VIRUP} as fast as possible. In particular, the nodes data layout is already the same as the layout \texttt{VIRUP} passes to the graphics driver. The nodes can then be uploaded directly from the disk to the GPU memory without further computation.

\subsubsection{Reference frame and spacial scales}

To account for the huge range of spatial scales covered by the data, up to 27 orders of magnitudes, 
while using single-precision variables, we change the reference frame of the leaf node 
to which the camera is the closest. 
Indeed, getting too close to a particular point will make it shake due to low precision of the 32-bits floating-point variables (about 7 digits of precision). When the camera enters a leaf node, all of its 16'000 data points are moved to correspond to coordinates of a reference frame which has its origin to the closest point to the camera. When this closest point changes, the coordinates of all the data points are re-computed to be centered on the new closest point.
This way, no matter how close to a point the camera gets, the coordinates of the camera relative to this point will remain close to zero, which behaves perfectly well in a floating-point representation.

\subsubsection{Volumetric rendering}

To render dust and gas from a galaxy, as in our Milky Way model for example, we compute 3D density maps 
of the absorbing media from the simulated snapshot. Those density maps will be loaded in GPU memory and
further used to perform ray-marching. 

In our galaxies, we consider stars as light sources.
In order to properly render a star including dust absorption, we first compute the luminance of
the corresponding point. As dust is only absorbing (no emission), we simply need to march from the 
light source down to the camera once to get the absorption coefficient and finally get the resulting 
luminance.
\texttt{VIRUP}  also includes emissive media, as for example the H-alpha emission of ionized gas.
Those media are ray-marched from the boundary of their maps to the camera, including
absorption, to obtain a total density value.
We ray-march both emissive and absorbing media at the same time, so that the absorption of emissive media is correctly computed along the depth.

\begin{figure}[h]
  \centering
  \includegraphics[width=0.48\textwidth]{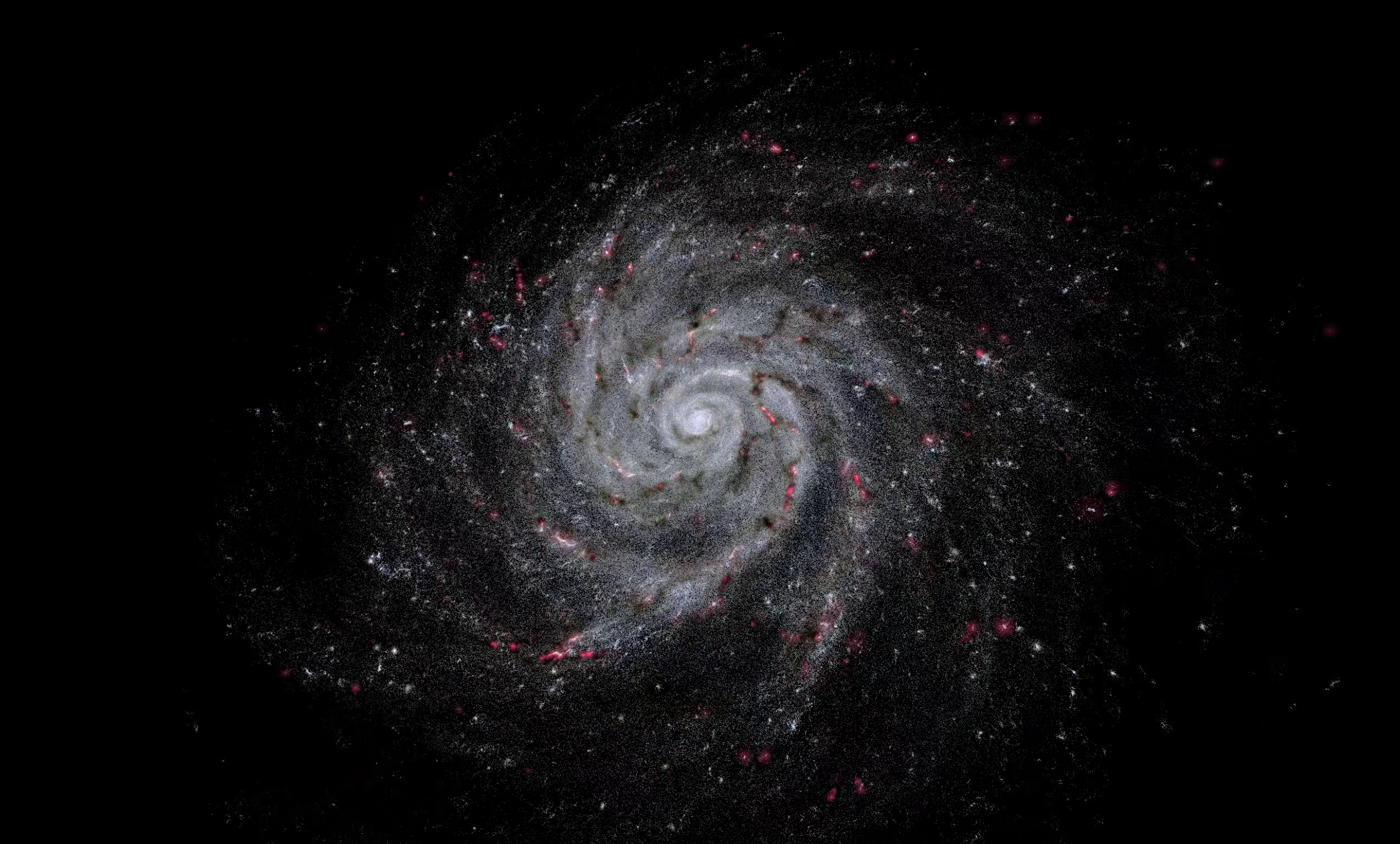}
  \caption{The \texttt{AGORA} Milky Way-like galaxy simulation 
  displayed with a volumetric rendering including H$-\alpha$ emission and dust absorption.}
\end{figure}

\subsubsection{Tone mapping}

The final state of \texttt{VIRUP}'s main framebuffer\footnote{The buffer that contains the pixels 
resulting from all the main drawing commands.} before post-processing is a luminance map in SI units. 
There is a subtlety concerning the rendering of points that is worth developing here,
as \texttt{VIRUP} renders a lot of point-based data. 
As points are objects without dimension, luminance is not defined. 
However, we can associate to each points its illuminance which depends solely on its apparent magnitude 
obtained from the luminosity of the object represented by the point and its distance to the camera.
This way, any illuminance associated to a point, that is written in the framebuffer 
is converted to a luminance by simply dividing it by the solid-angle of a pixel, ensuring
energy conservation.

An important step of the post-processing is the conversion of the luminance to an RGB value 
which will be displayed on the screen, a step also known as tone mapping.
We tried to be as realistic as possible matching what the human eye would see. 
To this end, we use the concepts of dynamic range and exposure.

The dynamic range is the factor between the dimmest visible luminance ($l_{\rm min}$) and the brightest one
($l_{\rm max}$), without considering pixel saturation. To roughly simulate a human eye, this factor must typically 
be set to $10'000$. 
We thus set $l_{\rm min}=1$ and $l_{\rm max}=10'000$. As typical LCD screens output pixel values from  
$0$ to $255$ in integer values,  we first need a function that maps the $[1, l_{\rm max}]$ interval 
(1 being the dimmest visible luminance) to  $[\frac{0.5}{255}, 1]$  (fragment shaders output in the $[0;1]$ range, 
which is then multiplied by 255). 
We use $0.5$ as the smallest visible pixel value as it will be rounded to 1 by the discretization. 
The mapping function also needs to be a logarithmic function if we want to see details at all scales, 
especially in the dimmest regime.

We used the following mapping function that fulfill the previous requirements:
\begin{equation}
    f(x) = \left(  
    \frac{\log\left(1+ (x-l_{\rm min})/c_2 \right)}{
          \log\left(1+ (l_{\rm max}-l_{\rm min})/c_2 \right)}     \right) 
          \left(1-c_1\right) + c_1
\end{equation}
where $c_1 = \frac{0.5}{255}$ and $c_2$ is a constant that defines the elbow of the logarithm in the range $[l_{\rm min}, l_{\rm max}]$ and
is set to about 1000.
Note that the factor 255 can easily be adapted to account for different screen dynamic ranges, useful for example
for HDR ports.

The exposure is a factor that will ensure that the dimmest visible luminance of the scene is 
mapped to 1 before being passed to our mapping function $f$. 
As a human eye can typically see in a range of luminance ranging from $10^{-6}$ to $10^8\,\rm{cd/m^2}$, 
we can restrict the exposure factor to the range  $[1/(10^8-l_{\rm max}),1/10^{-6}]$. 
Here, $10^8-l_{\rm max}$ corresponds to the dimmest visible luminance when the absolute visible brightest luminance is $10^8\,\rm{cd/m^2}$.

We added an optional auto-exposure function that controls the exposure factor over time depending on the scene 
average luminance, using a higher weight at the center of the scene. 
This function follows curves of the adaptation properties of a typical human eye that were tailored to 
match those found in \cite{dimitrov2008measuring}.

To get the final tone mapping function, we multiply our scene luminance by the exposure factor before passing it to our mapping function $f$ which outputs a normalized pixel value. 

For additional realism, we can also simulate the fact that human eyes can no longer see colors below a certain 
luminance threshold. This is called the Purkinje effect. We supplemented \texttt{VIRUP} with a simulation of this effect, which gradually transforms RGB luminances below a certain threshold to grayscale.

All these parameters : dynamic range, exposure and the toggleable Purkinje effect allow for 
a wide range of photo-realistic renderings. It is then possible to mimic either human eyes or photographic 
devices. As an example, using a low dynamic range will mimic the rendering of an old spacecraft camera.

\subsection{Planetary systems}


\subsubsection{Interpolation of Kepler Orbits}

To obtain the most accurate positions of the bodies of the Solar System at a given time, we use data 
from the NASA JPL Horizons tool. 
However, as \texttt{VIRUP} is designed to be able to run offline, we cannot
retrieve those positions from the NASA tool at launch time or when a new date is set.
Instead, we rely on a local database. For each celestial body, we have downloaded its 
known orbital parameters at variable rates,  daily for years close to 2021 (about 200 years before and after), 
weekly or more for the dates beyond this time range. Note that this data does not weight more than 
ten megabytes per body.
To get the position of a body at any specific time, the corresponding orbital parameters are interpolated linearly 
between the two closest dates of our database that encompass this time.
This is way more accurate than linearly interpolating position-vectors, and gives results precise enough for most cases.

The proper treatment of the orbits of space probes adds another difficulty. 
Indeed, they can strongly modify their orbits on time scales much shorter
than one day, the smallest default time sampling of our database. Such behaviour occurs typically when
the probe fired up its thrusters for example.
We thus detect such event and retrieve orbital parameters
at a much larger time resolution to ensure an optimal interpolation.

Peculiar cases exist when passing the elliptical/hyperbolic orbit transition. It happens when a probe gets captured by a planet or when it achieves escape velocity for example. There, 
the interpolation of the orbital parameters is no longer adequate as the 
semi-major axis goes from minus infinity to plus infinity or vice-versa.
In such cases, to keep it as accurate as possible, the orbit is  determined by a non-linear interpolation of the state-vectors. This interpolation is however performed only during the shortest possible time interval.

\subsubsection{Atmospheres}

Our atmospheres rendering is mostly based on \cite{hillaire2020scalable}. 
The main difference is that, as VIRUP's terrain rendering doesn't account for terrain height yet,
we do not need to use their Aerial Perspective LookUp Table (LUT) or their Multiple Scattering LUT.
Multiple scattering effects in atmospheric rendering are mostly visible in terrain-induced shadows. 
As our planets still have some ellipsoidal terrain,  aerial perspective effect must be  taken into account,
however, there is no need to include a LUT with several layers. 
We split the atmospheric rendering equation into two distinct parts : the in-scattering (or emissive) part and the out-scattering (or absorbing) part. 
Both components are stored in LUTs with similar parametrization as the Sky-View LUT of \cite{hillaire2020scalable}. 
Those LUTs can be computed without terrain data while still taking terrain into account, as the terrain is described
analytically as an ellipsoid.
\begin{figure}[h]
  \centering
  \includegraphics[width=0.48\textwidth]{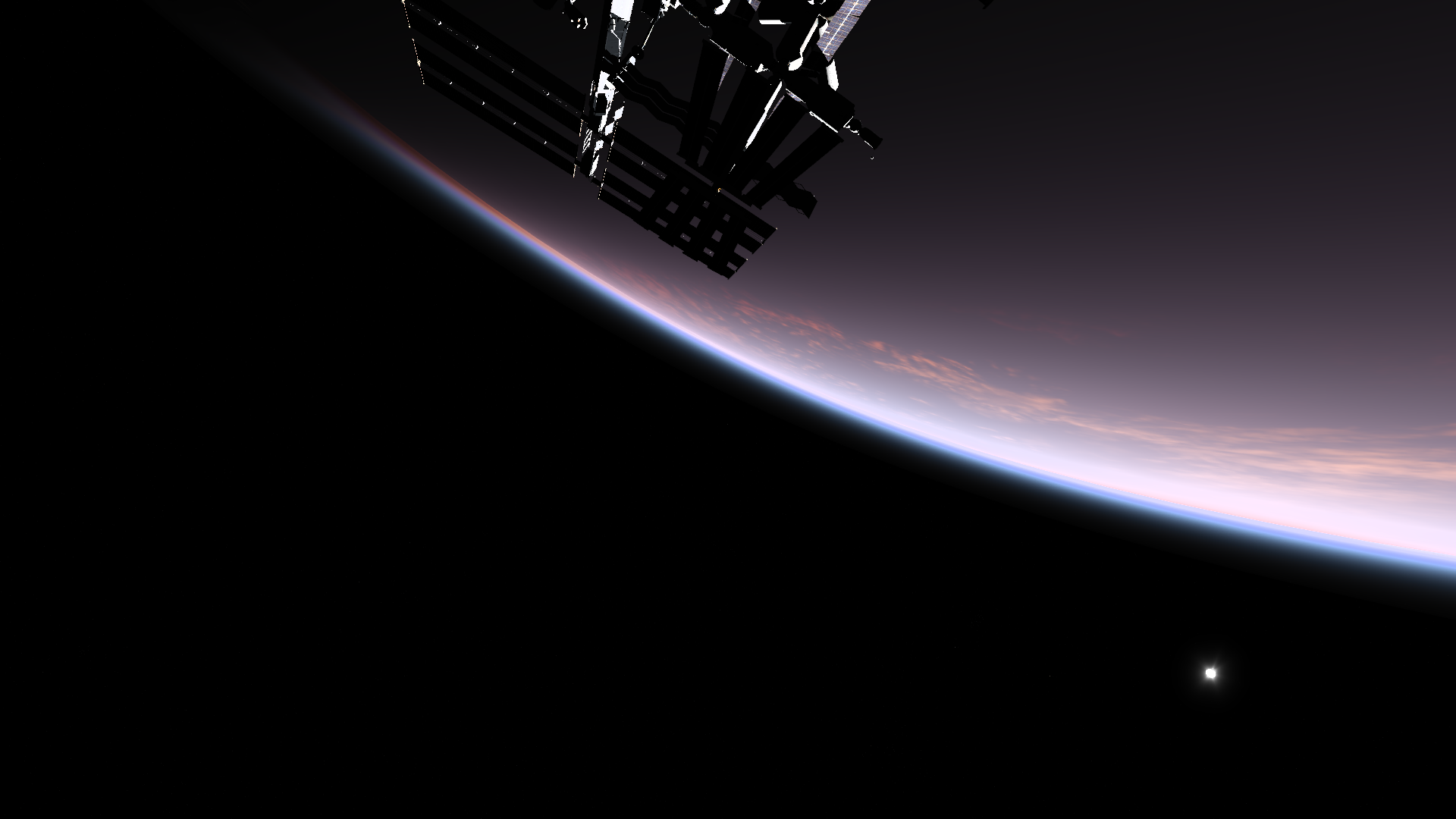}
  \caption{Earth's atmosphere as seen from the ISS near sunset.}
\end{figure}

\subsubsection{Celestial bodies light interaction}

Celestial bodies close to each other (mostly planet-moon systems or probes with other bodies) reflect light from 
their star on each other.  As the star is usually not a point-like source, they also cast smooth shadows on each other.
In computer graphics, a standard way to handle these subtle effects is to use global illumination algorithms or shadow
mapping. 
As most of the time, celestial bodies can be approximated by spheres, we can use a much more efficient method.

First, it is worth mentioning that while we take into account light and/or shadows cast on probes by planets or moons, 
we neglect the reverse, i.e., we neglect light and/or shadows cast by probes on planets or moons.
When an object reflects light on another, for example the Moon reflecting the Sun's light on  Earth, we consider the 
reflector to be a directional light source. We can compute the Moon's phase as seen from  Earth to get the amount 
of light reflected by the Moon and choose the Moon's direction as the light source direction towards the Earth. 
Note that we do not perform multiple light bounces as we do not consider that some Moon light gets reflected 
back from Earth to the Moon and so on. 
So, the Earth (resp. the Moon) reflects light from the Sun to the Moon (resp. the Earth) only once.
A difficulty appears however when moons are relatively close to their planet, as is Io to Jupiter, 
or the ISS to the Earth for example. 
In these situations, we cannot compute Earth's phase and 
direction to infinity as seen from the ISS, as the ISS would still receive light from the planet as it would 
be way past sunset. 
In these cases, we compute an approximation of a corrected phase corresponding to the light receiver field of 
view of the reflector and a corrected light direction that points more or less towards the barycenter of the light 
reflected by the reflector as seen by the receiver.

When an object casts a soft shadow on another, for example when a solar eclipse occurs and the Moon casts a shadow 
over Earth's surface, we consider the light source as a disks at infinity occluded by another disk at infinity. 
As bodies can be ellipsoids, we first perform a transformation into a space where the occluder is a sphere. 
This also transforms the light source into this space. 
While this is incorrect, it guarantee the shadow to have a correct shape. The error induced by this approximation
only affects the smoothness of the shadow. 

In the particular case where  both the light source and the 
occluder are spheres, both will be seen as disk from any 
point on the shadow-receiver's terrain or on the atmosphere. 
We then simply compute the ratio of light that still reaches this point to adjust its luminosity, 
which is a trivial geometric problem to solve.


%
\begin{figure}[h]
  \centering
  \includegraphics[width=0.48\textwidth]{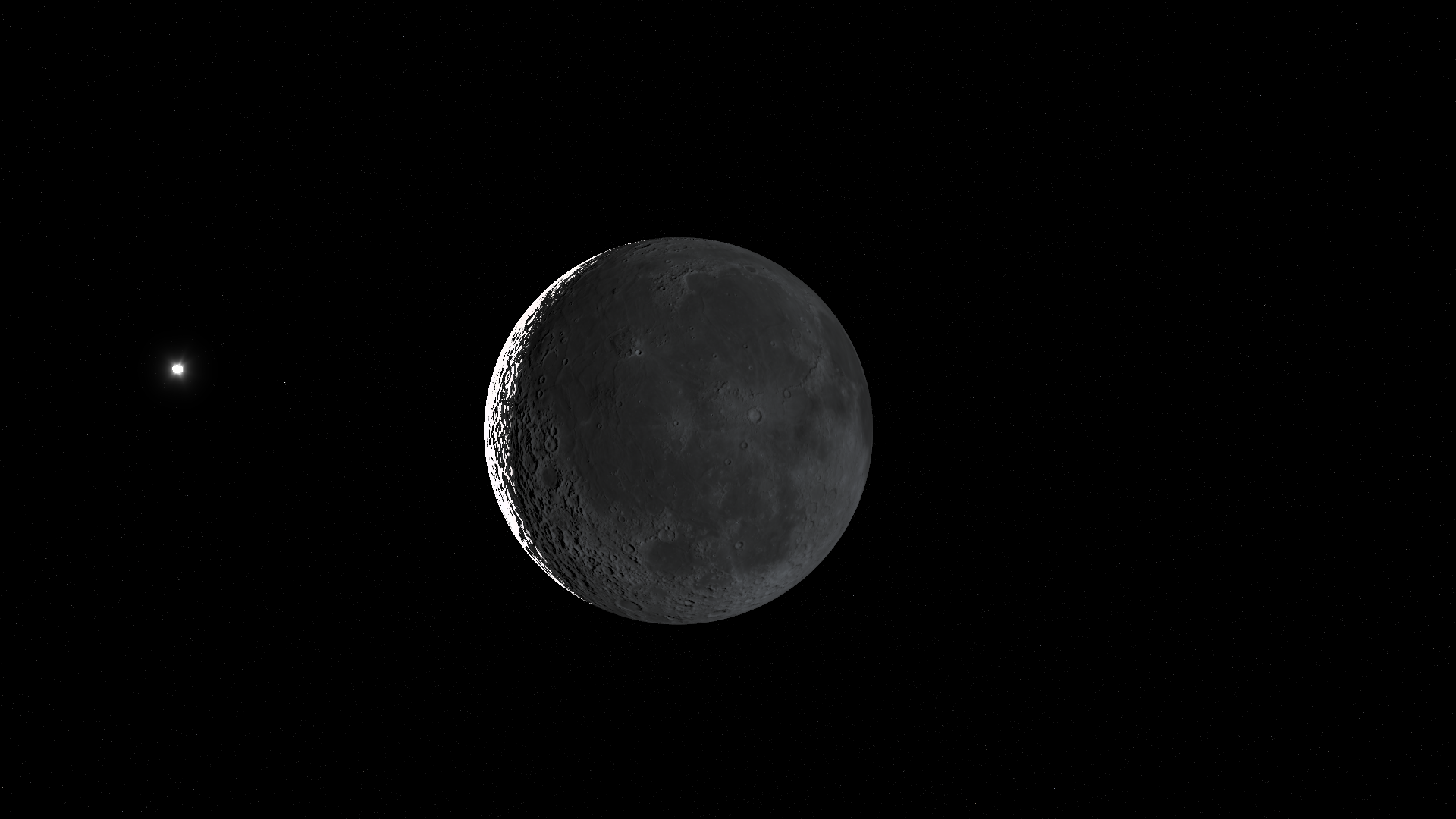}
  \caption{The Moon during a new moon, lit mostly by Earth's reflection of sunlight.}
\end{figure}
\begin{figure}[h]
  \centering
  \includegraphics[width=0.48\textwidth]{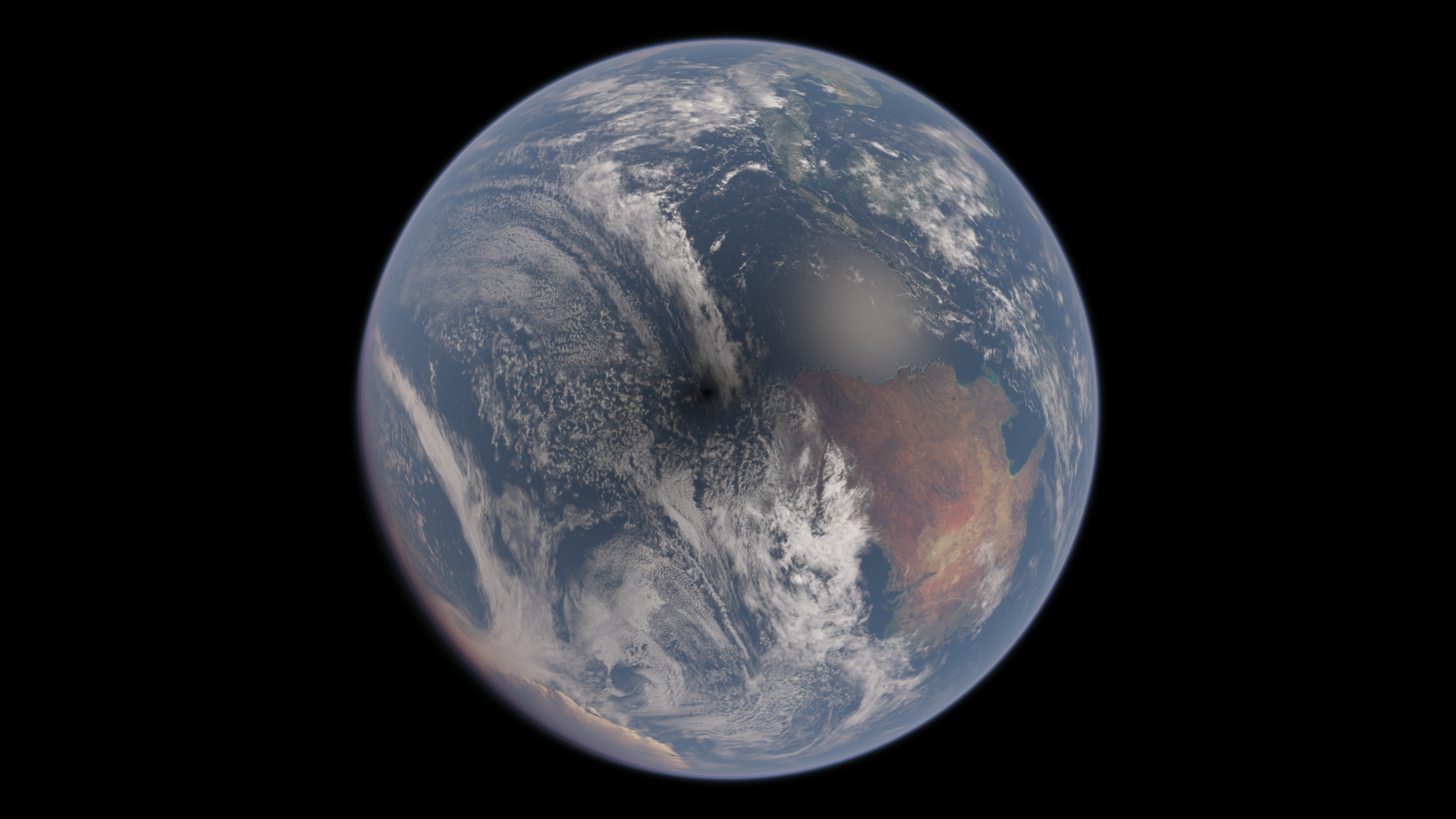}
  \caption{Solar eclipse that will occur on April 20th, 2023. We can see the Moon's soft shadow on the south-west coast of Australia.}
\end{figure}
%

%

%

\section*{Acknowledgements}

%

We would like to thank Lo\"ic Hausammann and Mladen Ivkovic for very useful discussions.  
\texttt{VIRUP} has been supported  by the Swiss National Science 
Foundation under the AGORA Grant CRAGP2\_178592.
It also received support from the EPFL Interdisciplinary Seed fund.

%
\bibliographystyle{aa}
\bibliography{bibliography}
%
\clearpage
\onecolumn

\end{document}